\documentstyle[epsfig,12pt,preprint,tighten,aps]{revtex}
\begin{document}
\draft
\title{
\rightline{{\tt Feb 2001}}
\rightline{{\tt UM-P-2001/005}}
\rightline{{\tt Astro-ph/0102294}}
\
\\ 
Seven (and a half) reasons to believe in Mirror Matter: From
neutrino puzzles to the inferred Dark matter in the Universe
}
\author{R. Foot}
\maketitle
\begin{center}
{\small \it School of Physics\\
Research Centre for High Energy Physics\\
The University of Melbourne Vic 3010
\\ Australia \\
foot@physics.unimelb.edu.au}\end{center}
\vspace{-0.5cm}
\begin{abstract}
Parity and time reversal are obvious and plausible candidates
for fundamental symmetries of nature.
Hypothesising that these symmetries exist implies
the existence of a new form of matter, called mirror matter. 
The mirror matter theory (or exact parity model) makes four 
main predictions: 
1) Dark matter in the form of mirror matter 
should exist in the Universe (i.e. mirror galaxies, stars, 
planets, meteoroids...),
2) Maximal ordinary neutrino - mirror neutrino oscillations 
if neutrinos have mass, 
3) Orthopositronium should have a shorter effective  
lifetime than predicted by QED (in ``vacuum" experiments)
because of the effects of photon-mirror photon mixing and 
4) Higgs production and decay rate should be $50\%$ lower than in 
the standard model due to Higgs mirror - Higgs mixing (assuming
that the seperation of the Higgs masses is larger than their decay widths).
At the present time there is strong experimental/observational
evidence supporting the first three of these predictions, while 
the fourth one is not tested yet because the Higgs boson,
predicted in the standard model of particle physics, is yet
to be found.  This experimental/observational evidence
is rich and varied ranging from the atmospheric and solar
neutrino deficits, MACHO gravitational microlensing events, 
strange properties of extra-solar planets, the existence 
of ``isolated" planets, orthopositronium lifetime anomaly, 
Tunguska and other strange ``meteor" events including perhaps, 
the origin of the moon.  The purpose of this article is to provide
a not too technical review of these ideas along with some 
new results.

\end{abstract}

\newpage
One thing that physicists have learned over the
years is that the interactions
of elementary particles obey a variety of symmetries. 
Some of these symmetries are quite familiar such as rotational 
invariance and translational invariance - physics text books are 
the same in Melbourne as they are in Moscow (once you
translate them!). There are also
other, less familiar symmetries such as
gauge invariance and (proper) Lorentz invariance, which
are nevertheless quite elegant and natural once you get
to know them.
Of course, it is pertinent to recall that the invariance
of particle interactions under these symmetries was
not always so obvious. For example, Dirac showed
us that the (quantum mechanical) interactions 
of the electron were only compatible with (proper) Lorentz invariance
if positrons (i.e. anti-matter) existed. Fortunately for
Dirac, his startling prediction was soon verified by
experiments.  

Remarkably though, experiments in the 1950's and 1960's
showed that space reflection symmetry (parity) and time
reflection symmetry (time reversal) do not appear 
to be fundamental symmetries of particle interactions.
For example, in well known beta decay processes, such
as $p \to n \ + \ e^+ \ + \ \nu_e$, the electron 
neutrinos\footnote{
The neutrino is a class of weakly interacting elementary 
particle with intrinsic spin 1/2. High Energy Physics 
experiments have revealed that
3 different ``species" of neutrino exist, called electron 
neutrinos ($\nu_e$), muon neutrinos ($\nu_\mu$) and
tau neutrinos ($\nu_\tau$).}, $\nu_e$ 
{\it always} have
their spin angular momentum aligned opposite to their
direction of motion (similar to a ``left handed" cork screw).
Nobody has ever observed a ``right handed" neutrino.
However, just as (proper) Lorentz invariance required
the existence of anti-matter, it turns out it is possible
for particle interactions 
to conserve also the improper Lorentz transformations
of parity and time  reversal if a new form of matter exists
- mirror matter.  In this theory\cite{flv},
each ordinary particle, such as the photon, electron, proton
and neutron, has a corresponding mirror particle, of 
exactly the same mass as the ordinary particle. 
The parity symmetry interchanges the ordinary particles with the
mirror particles [as well as $(x,y,z,t) \to (-x,-y,-z,t)$]
so that the properties of the mirror
particles completely mirror those of the ordinary particles
\footnote{
It also also possible to envisage variant theories for which
the symmetry is broken so that the mirror particles
have masses which are different from the ordinary particles.
Such theories though, are typically more complicated and less
predictive, as well as being less elegant. See Ref.\cite{jhep} for a 
discussion of these variants.  Also note that the mirror matter
model is also compatible with many extensions of the standard
model including: Grand Unification, Supersymmetry, Technicolour, 
Extra dimensions
(large and small), Superstring theory (especially
$E_8 \otimes E_8$) etc.}.
For example the mirror proton and mirror electron are stable and 
interact with the mirror photon in the same way in which the
ordinary proton and electron interacts with the ordinary photons.
The mirror particles are not produced
in Laboratory experiments just because they couple very
weakly to the ordinary particles. In the modern language of gauge
theories, the mirror particles are all singlets under 
the standard $G \equiv SU(3)\otimes SU(2)_L \otimes U(1)_Y$
gauge interactions. Instead the mirror
fermions interact with a set of mirror gauge particles,
so that the gauge symmetry of the theory is doubled,
i.e. $G \otimes G$ (the ordinary particles are, of 
course, singlets under the mirror gauge symmetry)\cite{flv}.
Parity is conserved because the mirror particles experience
right-handed mirror weak interactions
while the ordinary particles experience the usual 
left-handed weak interactions.  Ordinary and mirror
particles interact with each other predominately by gravity only.

While parity is obviously an extremely attractive theoretical candidate
for a symmetry of nature, its existence cannot, unfortunately, be proven
by pure thought (well at least nobody has done so 
up to now). Whether or not nature is left-right symmetric 
will be decided by experiments.  The mirror matter theory 
makes four main experimentally testable predictions:
\vskip 0.1cm
\begin{itemize}
\item
Mirror matter (e.g. mirror hydrogen composed of mirror protons
and mirror electrons) should exist in the Universe and would appear
to us as Dark Matter\cite{kb}.
Specifically, mirror galaxies, mirror stars, mirror planets 
and perhaps even mirror meteoroids could all exist.
\item
If neutrinos are massive (and non-degenerate) 
then oscillations between ordinary
and mirror neutrinos are maximal\cite{flv2,f}.
\item
Orthopositronium should have a shorter effective lifetime (in a ``vacuum"
experiment) than predicted in QED
due to the effects of photon - mirror photon kinetic mixing\cite{gl,fg}.
\item
Higgs production and decay rate should be $50\%$ lower than
in the standard model of particle physics due to Higgs - mirror Higgs 
mixing\cite{flv2,flv}.
This holds assuming that the two mass eigenstate Higgs fields 
have mass separation much larger than their decay widths
\footnote{
On the other hand, if the mass splitting is very small 
then there will be no experimentally observable mirror Higgs effect 
(see Ref.\cite{ben} for a detailed study).}.
\end{itemize}

\noindent
At the present time there is strong experimental/observational
evidence supporting
the first three of these predictions, while the fourth one is not
tested yet because the Higgs boson, predicted in the standard
model of particle physics, has yet to be found (though it may be found
in collider experiments in the near future).
This experimental/observational evidence can
be viewed as an explanation to seven specific scientific puzzles 
(most of them long standing), which we list below:
\vskip 0.2cm
\noindent
1) Massive Astrophysical Compact Halo Objects (MACHOs):
Invisible star-sized objects in the halo
of our galaxy identified by their gravitational effects 
in microlensing searches. 
\vskip 0.2cm
\noindent
2) Close-in extrasolar planets: Large gas giants only
$\sim 7 $ million kilometers ($0.05$ A. U.) from their star,
where it is too hot for them to form. 
\vskip 0.2cm
\noindent
3) Recent discovery of ``Isolated planets" in the Sigma
Orionis star cluster:
The properties of these objects are unexplained by existing theories.
\vskip 0.2cm
\noindent
4) Solar neutrino deficit: Half of the electron neutrinos emitted
by nuclear reactions in the solar core are missing.
\vskip 0.2cm
\noindent
5) Atmospheric neutrino deficit: Half of the 
(up-going) $\nu_\mu$ produced
as a consequence of cosmic ray interactions with the atmosphere
are missing.
\vskip 0.2cm
\noindent
6) Orthopositronium lifetime anomaly: 
A precision vacuum cavity experiment finds a lifetime shorter than 
the standard model prediction.
\vskip 0.2cm
\noindent
7) Disappearing meteors: 
Tunguska (and Tunguska-like events) including, perhaps,
the origin of the moon.

We now describe how the mirror matter theory explains 
these 7 scientific puzzles. 

\vskip 0.3cm
\noindent
{\bf 1), 2) \& 3) Implications of the mirror world for cosmology:
MACHOs, Extra-solar planets \& ``isolated planets"}
\vskip 0.3cm
\noindent
There is strong evidence for a large amount 
of exotic dark matter in the Universe (maybe as much 
as 95\% of the mass of the Universe). For example,
the orbits of stars at the (visible) edge of our galaxy
provide information about the distribution of matter
within our galaxy. These observations show that
there must exist invisible halos in
galaxies such as our own Milky Way. Furthermore, there is also strong
evidence that this dark matter must be something exotic:
ordinary matter simply cannot account for it\cite{freeze}.
Mirror matter is naturally dark (because the coupling 
of mirror matter to ordinary photons is necessarily
very small\footnote{
A very small coupling between ordinary and mirror photons 
is allowed by the theory
and is suggested by an experiment measuring the orthopositronium
lifetime (see next section) but this interaction
is too small to make mirror matter directly observable\cite{ig}.})
and is a very
natural candidate for the inferred dark matter in the Universe.
This has been argued for some time by Blinnikov and
Khlopov\cite{kb} (see
also the recent reviews in \cite{bd}).
The physics of galaxy formation in the early Universe is
far from being completely understood. Phenomenologically, one envisages
galaxies containing some mixture of ordinary and mirror 
matter. In fact just about anything is possible:
Galaxies ranging from no mirror matter, to galaxies composed
almost entirely of mirror matter.
Mirror matter inside galaxies will fragment into diffuse
clouds and eventually into mirror stars. This type of collapse
should happen quite independently for ordinary and mirror
matter, since they will have locally different
initial conditions such as angular momentum and abundance
as well as chemical composition\footnote{ 
For example, the rate of collapse of mirror matter in a 
diffuse gas cloud will obviously 
occur at a different rate to ordinary matter in the cloud
because collapse requires non-gravitational
dissipative processes (such as atomic collisions) to release 
the energy so that the system can become more tightly bound. 
These dissipative processes will occur at different rates for ordinary
and mirror matter due to their different initial conditions
and chemical composition.}. 
Thus, dark matter made of mirror matter would have the property of
clumping into compact bodies such as mirror stars, however
their distribution within the galaxy can be
quite independent of the distribution of ordinary matter.
Dark matter composed of mirror matter
thus leads naturally to an explanation\cite{exp} for the mysterious
massive astrophysical compact halo objects (or MACHO's)
inferred by the MACHO collaboration\cite{macho}.
This collaboration has been studying the nature of halo dark 
matter by using the gravitational microlensing technique.
This Australian-American experiment has so far collected 5.7 years
of data and has provided statistically strong evidence for
dark matter in the form of invisible star sized objects which
is what you would expect if there was a significant
amount of mirror matter in our galaxy\cite{exp}.
The MACHO collaboration\cite{macho} 
have done a maximum likelihood analysis which implies
a MACHO halo fraction of $20\%$ for a typical halo model
with a $95\%$ confidence interval of $8\%$ to $50\%$.
Their most likely MACHO mass is between $0.15 M_{\odot}$ and
$0.9M_{\odot}$ depending on the halo model.
These observations are consistent with a mirror matter halo because
the entire halo would not be expected to be
in the form of mirror stars. Mirror gas and dust would
also be expected because they are a necessary consequence
of stellar evolution and should therefore significantly
populate the halo. 

If mirror matter does indeed exist in our galaxy, then
binary systems consisting of ordinary and mirror matter
should also exist. While systems containing approximately
equal amounts of ordinary and mirror matter are unlikely due
to the differing rates of collapse for
ordinary and mirror matter (leading to a local segregation of
ordinary and mirror matter), systems containing predominately ordinary 
matter with a small amount of mirror matter (and vice versa)
should exist. Interestingly, there is remarkable evidence for
the existence of such systems coming from extra-solar planet astronomy.

In the past few years more than 50 ``extrasolar" planets
have been discovered orbiting nearby stars\cite{www}. They 
reveal their presence because their gravity tugs
periodically on their parent stars leading to
observable Doppler shifts. In one case, the planet HD209458b,
has been observed to transit its star\cite{trans} allowing
for an accurate determination of the size and mass for this system.
One of the surprising characteristics of the extrasolar planets
is that there are a class of large ($\sim M_{Jupiter}$) 
close-in planets with a typical orbital radius of 
$\sim 0.05 \ A. U.$, that is, about 8 times closer than
the orbital radius of Mercury (so called ``51 Pegasi-like"
planets after the first such discovery\cite{mq}). Ordinary
(gas giant) planets are not expected to form close to stars because the 
high temperatures do not allow them to form.
Theories have been invented where they form far from the
star where the temperature is much lower,
and migrate towards the star. While such theories are possible,
there are also difficulties, e.g. the recent discovery
of a close-in pair of resonant planets\cite{marcy} is unexpected since
migration tends to make the separation between planets diverge
(as the migration speeds up as the planet becomes closer to the star).

A fascinating alternative possibility presents itself in the mirror
world hypothesis. The close-in extrasolar planets may be mirror worlds
composed predominately of mirror matter\cite{plan}. They do not
migrate significantly, but actually formed close to the star
which is not a problem for mirror worlds because
they are not significantly heated by the radiation
from the star. This hypothesis can explain the
opacity of the transiting planet HD209458b because mirror
worlds would accrete ordinary matter from the solar
wind which accumulates in the gravitational potential
of the mirror world\cite{footr}. It turns out that the effective
radius of ordinary matter depends relatively sensitively on the
mass of the planet, so that this mirror world hypothesis
can be tested when more transiting planets are discovered\cite{footr}.

If this mirror world interpretation of the close-in extrasolar planets
is correct then it is very natural that the dynamical
mirror image system of a mirror star with an ordinary planet
will also exist. Such a system would appear to ordinary observers
as an ``isolated" ordinary planet. Remarkably,
such ``isolated" planets have recently been identified
in the $\sigma$ Orionis star cluster\cite{iso}.
These planets have estimated mass of $5-15 M_{Jupiter}$
(planets lighter than this mass range would be too faint to
have been detected in Ref.\cite{iso})
and appear to be gas giants which do not seem
to be associated with any visible star.
Given that the $\sigma$ Orionis cluster is estimated to be less than  
5 million years old, the formation of these ``isolated" planets
must have occurred within this time (which
means they can't orbit faint stellar bodies such
as white dwarfs). Zapatero Osorio et al\cite{iso}
argue that these findings pose a challenge to
conventional theories of planet formation which are
unable to explain the existence of numerous isolated planetary mass
objects. Thus the existence of these planets is very surprising 
if they are made of ordinary matter,
however there existence is quite natural from the mirror world
perspective\cite{iso2}. Furthermore, if the isolated planets
are not isolated but orbit mirror stars then
there must exist a periodic Doppler shift detectable
on the spectral lines from these planets. This
represents a simple way of testing this hypothesis\cite{iso2}.

There is also recent, tantalizing observational evidence
for mirror matter from another source: A recent
weak gravitational microlensing study\cite{Erben}
has apparently discovered an invisible dark concentration of 
mass in the vicinity of the cluster, Abell 1942.
A fascinating possibility is that a mirror
galaxy (or galaxy cluster) containing virtually no 
ordinary matter has been discovered.
Further studies (such as Ref.\cite{gray}) should
help clarify whether this mirror matter interpretation 
is correct.

Finally, let us also mention that the existence of mirror matter 
may have interesting consequences for early Universe cosmology. 
However, early Universe cosmology is not precise enough yet to 
shed much light on mirror matter (although forthcoming precision 
measurements of the cosmic microwave background may help). 
For some recent articles on the implications of mirror matter 
for early Universe cosmology, see Ref.\cite{ec}.

\vskip 0.3cm
\noindent
{\bf 4) \& 5) 
Implications of the mirror world for neutrino physics:
Solar and atmospheric neutrino deficits}
\vskip 0.3cm
\noindent
It was realized in 1991\cite{flv2} and further studied in
Ref.\cite{f},
that neutrino oscillations of ordinary neutrinos
into mirror neutrinos would provide a simple way of testing the
mirror world hypothesis.
Neutrino oscillations are a well known quantum mechanical effect
which arise when the flavour eigenstates are linear combinations
of 2 or more mass eigenstates. For example, if the
electron and muon neutrinos have mass which mixes
the flavour eigenstates, then in general the
weak eigenstates are orthogonal combinations of mass
eigenstates, i.e.
\begin{equation}
\nu_e = \sin\theta \nu_1 + \cos\theta \nu_2,\
\nu_\mu = \cos\theta \nu_1 - \sin\theta \nu_2.
\label{1}
\end{equation}
A standard result 
%(see e.g. Perkins\cite{per}) 
is that the oscillation
probability for a neutrino of energy $E$ is then
\begin{equation}
P(\nu_\mu \to \nu_e) = \sin^2 2\theta \sin^2 L/L_{osc}, 
\label{osc}
\end{equation}
where $L$ is the distance from the source and $L_{osc} \equiv 4E/\delta
m^2$ is the oscillation length (and natural units
have been used, i.e. $c=h/2\pi=1$)\footnote{
In Eq.(\ref{osc}), $\delta m^2 \equiv m_1^2 - m_2^2$ 
is the difference in squared
masses of the neutrino mass eigenstates.}.
If $\sin^2 2\theta = 1$ then the oscillations have the greatest
effect and this is called maximal oscillations.
In our 1992 paper we found the remarkable result that
the oscillations between ordinary and mirror
neutrinos are necessarily maximal which is a direct consequence
of the parity symmetry.
One way to see this is to note that if neutrinos mix then the
mass eigenstates are non-degenerate and 
necessarily parity eigenstates if
parity is unbroken. Considering the electron neutrino,
$\nu_e$ and its mirror partner, $\nu'_e$, 
the parity eigenstates are
simply $\nu^{\pm} = (\nu_e \pm \nu'_e)/\sqrt{2}$ (since parity
interchanges the ordinary and the mirror particles) and hence
\begin{equation}
\nu_e = {\nu^+ + \nu^- \over \sqrt{2}},\
\nu'_e = {\nu^+ - \nu^- \over \sqrt{2}}.
\end{equation}
Comparing this with Eq.(\ref{1}) we see that
$\theta = \pi/4$ i.e. $\sin2\theta = 1$ and hence maximal mixing!
Thus if neutrinos and mirror neutrinos have mass and mix together then
the oscillations between the ordinary and mirror neutrinos are
necessarily maximal. This simple observation nicely explains 
the solar neutrino 
deficit since the oscillations between $\nu_e \to \nu'_e$
reduce the flux of $\nu_e$ from the sun by a predicted 50\%
after averaging over energy and distance (provided of course that
the oscillation length is less than the distance between the earth and
the sun, which means $\delta m^2 \stackrel{<}{\sim} 3 \times 
10^{-10}\ eV^2$)\footnote{
In principle it is necessary to take into account matter
effects for neutrino propagation in the Sun and the Earth.
However the net effect is only a slight modification
to the naive 50\% $\nu_e$ flux reduction expected
for vacuum oscillations (which may 
nevertheless be important in some circumstances)\cite{crock}.}.  
Electron neutrinos emitted from the sun
arise from various nuclear reactions in the solar core. 
Theoretically the most important are the pp reaction chain
where two protons fuse together to form deuterium:
$p+p \to \ ^2H + e^+ + \nu_e$.  This neutrino flux can be 
most reliably predicted since it is directly related to the luminosity
of the sun.
There are 3 experiments specifically designed to measure the pp
neutrinos which are called SAGE, GALLEX and GNO. The SAGE and
GALLEX experiments began
running around 1991 with GNO starting in 1998.
Their most recent results normalized to
the theoretical prediction\cite{bah} are \cite{sage}
\begin{eqnarray}
0.52 &\pm & 0.06(exp) \pm 0.05 (theory)\ (SAGE),\nonumber \\ 
0.59 &\pm & 0.06 (exp) \pm 0.05 (theory)\ (GALLEX),\nonumber  \\
0.50 &\pm & 0.09 (exp) \pm 0.05 (theory)\ (GNO),
\end{eqnarray}
where the errors are $1-sigma$.
These results are consistent with the mirror matter prediction of
$0.50$.  More recently the Superkamiokande Collaboration have 
reported an energy independent (within
errors) recoil electron energy spectrum 
in their experiment designed to measure the $^8 B$ neutrinos
(i.e. neutrinos from the nuclear reaction, $^8B + e^- 
\to \ ^8_4Be + \nu_e$) , again finding only $50\%$ of the expected
solar flux. Again these results were predicted in
the mirror matter model\cite{flv2,f}.

During 1993 I first became aware of the atmospheric neutrino anomaly and
immediately recognized that this could be further important evidence
for mirror matter \cite{f}. This anomaly suggests that the muon neutrino
($\nu_\mu$) oscillates into some other neutrino species 
with large mixing angle (which was only weakly 
constrained in 1993). This anomaly can easily be
explained by the Mirror Matter model since, as we have discussed above,
it predicts that each of the known
neutrinos oscillates maximally with its mirror partner 
if neutrinos have mass. Thus in this case it is theoretically
very natural to explain the atmospheric neutrino anomaly via
$\nu_\mu \to \nu'_\mu$ oscillations (where $\nu'_\mu$ is the
mirror muon neutrino)\footnote{
Of course the main alternative case of $\nu_\mu \to \nu_\tau$
oscillations is also possible within this framework\cite{yoon},
but it doesn't seem to be quite so elegant.}.
With the new results from the superKamiokande experiment the 
prediction of maximal mixing has been confirmed with the 90\% 
allowed region\cite{sk}
\begin{equation}
0.85 \stackrel{<}{\sim}
\sin^2 2\theta \stackrel{<}{\sim} 1.0
\end{equation}
This is in nice agreement with the 1993 mirror matter prediction of 
$\sin^2 2\theta = 1$.

Of course neutrino oscillations into the mirror world is not the 
only possible solution to the neutrino anomalies. 
However it does provide an 
elegant explanation for the inferred maximal neutrino oscillations
of $\nu_e$ (solar) and $\nu_\mu$ (atmospheric)
which is good reason to take it seriously\footnote{
These days
it is often argued that\cite{skr} the solution to the atmospheric neutrino 
anomaly is $\nu_\mu \to \nu_\tau$ oscillations (and it may be), 
however the superKamiokande data itself cannot
yet distinguish the simplest mirror world explanation from
$\nu_\mu \to \nu_\tau$ oscillations\cite{foot2000}.}.
Furthermore, it is one of the few solutions to the
solar and atmospheric neutrino puzzles which is also
consistent with the LSND accelerator experiment\cite{lsnd}.
The LSND experiment provides strong evidence that mixing between
at least the first two generations is small (which is already known to happen
for quark mixing). 
Most importantly, the mirror world explanation will be
tested more stringently in the
near future from a variety of new experiments including
Borexino\cite{bor}, Kamland\cite{kam} and especially the
Sudbury Neutrino Observatory (SNO)\cite{sno}.
The latter experiment will be crucial in distinguishing
$\nu_e \to \nu'_e$ oscillation solution to the solar neutrino problem
with many other proposals, while
Borexino and Kamland will be very important in pinning down
the $\delta m^2, \sin^2 2\theta$ parameters (and
in the process stringently checking the mirror world prediction
of $\sin 2\theta = 1$).

\vskip 0.3cm
\noindent
{\bf 6) Implications of the mirror world for Laboratory
experiments: Orthopositronium lifetime anomaly}
\vskip 0.3cm
\noindent
There are essentially only 3 ways in which ordinary and mirror matter
can interact with each other besides gravity:
That is by photon mirror photon kinetic mixing\cite{hol,gl,flv}, 
Higgs - mirror Higgs interactions\cite{flv,flv2}, 
and by neutrino - mirror neutrino 
mass mixing (if neutrinos have mass)\cite{flv2,f}\footnote{
Only neutral particles can mix with each other since
electric charge is conserved. Mixing of say an electron
with a mirror electron would violate electric charge
conservation and such mixing is constrained to be negligible.}.

The effect of neutrino - mirror neutrino mass mixing
has already been described: It leads to maximal 
ordinary - mirror neutrino oscillations which
can simply and predictively
explain the atmospheric and solar neutrino deficits.
The effect of Higgs - mirror Higgs interactions
is to reduce the production and
decay rate by 50\% compared with the standard
model Higgs particle (provided that the Higgs
mass splitting is large enough). This prediction will be tested
when the Higgs is discovered which 
may occur soon either at Fermilab or at the Large
Hadron Collider at CERN. Finally, we have photon - mirror
photon kinetic mixing which leads to interesting
effects for orthopositronium.

Photon - mirror photon kinetic mixing
is described in quantum field theory as the Lagrangian term:
\begin{equation}
{\cal L}_{int} = {\epsilon \over 2}F^{\mu \nu} F'_{\mu \nu}
\label{km}
\end{equation}
where $F^{\mu \nu} \equiv \partial^{\mu} A^{\nu} - \partial^{\nu}
A^{\mu}$ is the usual Field strength tensor, and the $F'$ is
the corresponding quantity for mirror photons.
This Lagrangian term may be considered as a fundamental
interaction of nature\cite{flv} or may arise as
a quantum mechanical ``radiative correction" effect
\cite{hol} (see also Ref.\cite{cf}).
Glashow\cite{gl} has shown that the kinetic 
mixing term leads to a modification of the 
orthopositronium lifetime (which turns out to be the most
important experimental implication of photon - mirror photon 
kinetic mixing).  Recall orthopositronium is the bound state
composed of an electron and positron where the spins of
both particles are aligned so that the bound state has spin 1.
The ground state of orthopositronium (o-Ps) decays predominately into
3 photons. The decay rate has been computed in QED leading
to a discrepancy with some of the experimental measurements.
Some of the measurements find a faster decay rate than theoretically
predicted.  This discrepancy has lead to a number of experimental
searches for exotic decay modes, including a stringent limit
on invisible decay modes\cite{limit}.

The modification of the lifetime predicted in the mirror
matter theory occurs because the kinetic mixing of the photon
with the mirror photon generates a small off-diagonal
orthopositronium mass leading to oscillations between
orthopositronium and mirror orthopositronium.
The orthopositronium produced in the experiment oscillates
into its mirror partner, whose decays into three mirror photons
are undetected. This effect only occurs in a vacuum experiment
where collisions of the orthopositronium with background
particles can be neglected\cite{gn}. Collisions with background
particles will destroy the quantum coherence necessary
for oscillations to occur. Thus, experiments with large
collision rates remain unaffected by kinetic mixing
and the lifetime of orthopositronium will be the same
as predicted by QED. Experiments in vacuum on the other hand,
should show a slight increase in the decay rate,
as oscillations into mirror orthopositronium and their subsequent
invisible decays effectively reduce the number of orthopositronium
states faster than QED predicts.  The two most accurate experimental 
results, normalized to the theoretical QED prediction\cite{afs} are 
given in the table below 
\footnote{A third experiment with gas\cite{aa2} also has
an anomalously high decay rate, however there appears to be
large possible systematic uncertainties because their
are indications that the orthopositronium
may not be thermalized (as assumed) in this 
experiment\cite{therm,tok}.}
\vskip 0.7cm

{\begin{center}
\begin{tabular}{|l|l|l|l|}
\hline
Reference$\;\;\;\;\;\;\;$
&$\Gamma_{oPs}(exp)/\Gamma_{oPs}(theory)$ $\;\;\;$ 
&Method$\;\;\;\;\;\;\;\;\;\;\;\;$
&$\Gamma_{coll}$$\;\;\;\;\;\;\;\;\;\;\;\;$\\
\hline
Ann Arbor\cite{aa1}&$1.0012\pm 0.0002$&Vacuum Cavity&$\sim (3-10)\Gamma_{oPs}$\\
%Ann Arbor\cite{aa2}&$7.0514\pm 0.0014$&Gas&$\sim 10^3 \Gamma_{oPs}$\\
Tokyo\cite{tok}&$1.0000 \pm 0.0004$&Powder&$\sim 10^4 \Gamma_{oPs}$\\
\hline
\end{tabular}\end{center}}
\vskip 0.12cm
\noindent
Thus, we see that the Tokyo experiment agrees with the QED prediction
while the Ann Arbor vacuum experiment disagrees by about 6 sigma.
These results can be explained in the mirror matter model
by observing that the large collision rate ($\Gamma_{coll}$)
of the orthopositronium in the Tokyo experiment will
render oscillations of orthopositronium with its mirror
counterpart ineffective\footnote{
The experimental limit\cite{limit} for invisible decay modes
also does not exclude this mirror world oscillation
mechanism because the collision rate of the 
orthopositronium was very high in those experiments.}, 
while the larger decay rate obtained in the vacuum cavity
experiment can be explained
because of the much lower collision rate of orthopositronium
in this experiment allows the
oscillations of ordinary to mirror orthopositronium to take 
effect. The fit of the theory to the cavity
experiment implies that the kinetic mixing parameter is
$\epsilon \approx 10^{-6}$\cite{fg}.

While the mirror world can nicely explain
the orthopositronium lifetime puzzle, this puzzle
is based only on one anomalous vacuum cavity experiment
(however the statistical significance is impressive: 6 sigma).
Also, the value for $\epsilon$ is a bit too large
to be acceptable for early Universe cosmology
(BBN) unless of course, one of the standard assumptions
is wrong (which is certainly possible of course)\footnote{  
Consistency with
standard big bang nucleosynthesis (BBN) suggests that
$\epsilon \stackrel{<}{\sim} 3 \times 10^{-8}$\cite{cg}.
However, the cosmological situation is by no means clear.
For example, there are tentative indications from recent
precision measurements of the cosmic microwave background
that the energy density of the early Universe could be
about a factor of two larger than expected given the standard
particles (which is what
you would expect if $\epsilon \simeq 10^{-6}$ because
the effects of the photon - mirror photon kinetic mixing
interaction would then fully populate the mirror sector
in the early Universe). Also there may exist
a pre-existing or neutrino oscillation generated neutrino
asymmetry which may further modify things\cite{barifoot}.}. 
Clearly, what is really needed is a new experiment to check the 
anomalous vacuum cavity result.
In fact, an experiment with a larger cavity should make
things very clear, since there should be an even
larger mirror world effect if $\epsilon \approx 10^{-6}$.
Such an experiment has been
proposed to test for this effect and to confirm (or
reject) the mirror world explanation for the
orthopositronium lifetime anomaly\cite{ns}.

\vskip 0.3cm
\noindent
{\bf 7) Disappearing meteors: 
Tunguska (and Tunguska-like events) including, perhaps,
the origin of the moon.}
\vskip 0.3cm
\noindent

To summarise the current situation,  mirror matter is predicted to
exist if nature is left-right symmetric (i.e. parity invariant).
There is now considerable experimental/observational support 
for mirror matter coming from neutrino physics, the orthopositronium
lifetime puzzle, and astrophysics/cosmology.
It should be clear that further work in these fields
can really put the predictions of the mirror matter
to more stringent experimental/observational test. 

Given the simplicity and appeal of the mirror matter
theory and the large amount of experimental
evidence in favour of it, it is tempting
to entertain more fascinating (but also more speculative)
implications of the theory.
If mirror matter exists then perhaps one of the most
fascinating possibilities is that there is 
significant (by this I simply mean enough
to be ``observable") amount of mirror matter in our solar system.
While much of any initial mirror matter component in our solar
system may have found its way into the center of the sun\footnote{
Some may also be in the center of planets, but not so much. E.g.
one can deduce an upper bound of about $10^{-3}$ for
the fraction of mirror matter content of the Earth\cite{iv}.}
%assuming that the photon - mirror photon kinetic mixing
%can be neglected. 
%It is possible that this bound can
%be considerably relaxed for significant photon kinetic 
%mixing, such
%as the value suggested by the orthopositronium vacuum 
%cavity experiment where $\epsilon \simeq 10^{-6}$.}
(where its effects are more difficult to be unambiguously
observed) it is nevertheless possible for 
small mirror space bodies  (such as mirror meteoroids or mirror 
comets etc) could exist 
\footnote{It is also possible
that a large mirror body such as a mirror planet/star might
exist in our solar system if it is a relatively distant companion to
the sun\cite{sil,fs}.}. 
Such mirror bodies may not orbit in the same
plane as the ecliptic - they may orbit in a different
plane, or may be even spherically distributed (like the Oort cloud).
%\footnote{
%It is also possible that the mirror space bodies
%are associated with the hypothesised companion
%star to our sun (dubbed ``nemesis"), whose existence 
%has been suggested by the suspected periodic mass
%extinctions in the fossil record. Nemesis may be
%a mirror star which may explain why it has so far
%escaped detection\cite{sil}.}.

If such mirror bodies exist and happen to collide
with the Earth, what would be the consequences?
If the only force connecting mirror matter
with ordinary matter is gravity, then the consequences
would be minimal. The mirror space body would simply pass
through the Earth and nobody would know about it unless
the body was so heavy as to gravitationally affect the motion
of the Earth. However if there is kinetic
mixing between ordinary and mirror photons (which is suggested
by the orthopositronium experiment), then
the mirror space body would heat up as the nuclei of
the mirror atoms undergo Rutherford scattering
as they weakly interact [made possible because of the
photon - mirror photon kinetic mixing interaction,
Eq.(\ref{km})] with the nuclei of the ordinary oxygen and nitrogen
in the atmosphere.  The ordinary matter
which passes through the mirror meteoroid would also
heat up as the mirror meteoroid moves through the Earth's atmosphere.
This may make the mirror meteoroid effectively visible as it plumets
to the surface of our planet. 
There are essentially two possibilities (depending
on the chemical composition of the mirror meteoroid and 
also on the kinetic mixing parameter $\epsilon$): Either it disintegrates
in the atmosphere or it survives to reach the Earth's surface.
If it disintegrates no fragments will be found, since
mirror matter would be undetectable 
in our ordinary matter surroundings. 
If it survives and enters the ground then two things can
happen depending on the stopping distance ($D$). 
Either it is stopped over a short distance ($\stackrel{<}{\sim}
100$  meters)  
in which case the energy of the impact should leave a crater,
while if it is stopped over a large distance ($\stackrel{>}{\sim}$
few kilometers)
no impact crater would form since the meteoroid's kinetic
energy would be distributed over a large distance.
Again, in either case no meteoroid fragments would be found. 
It is straightforward to roughly estimate the stopping
distance $D$, which depends
on the strength of the kinetic mixing parameter, $\epsilon$,
the initial velocity of the space body ($v_i$), and also 
(more weakly) on the chemical composition of the
mirror space body and the density and composition of the
ground where it enters the Earths surface.
A rough calculation of the stopping distance in 
the Earth's crust in the case
of very small $\epsilon \stackrel{<}{\sim} 10^{-8}$
(using the surface density
of the Earth of $\rho \approx 3 \ g/cm^3$) gives
\footnote{
Note that the derivation of this equation will be published
at some point.}
\begin{equation}
D \sim \left({v_i \over 30\ km/s}\right)^4 \left({10^{-9} \over \epsilon}
\right)^2
 \ km.
\end{equation}
Thus, for $\epsilon \stackrel{<}{\sim} 10^{-9}$,
the typical stopping distance in the Earth's crust
is greater than about a kilometer.
Thus, such a body would {\it not} be expected to leave a 
large crater while for much larger values of $\epsilon$,
such as the value suggested by the orthopositronium 
experiment ($\epsilon \simeq 10^{-6}$),
the mirror meteoroid would release most of its
kinetic energy in the atmosphere, leading perhaps to an atmospheric
explosion. Remarkably there appears to
be significant evidence for ``disappearing meteors"\cite{olk},
i.e. meteors which are seen but do not lead to any impact crater and
no meteor fragments are found.
The most famous such event is the 1908 Siberian 
explosion (the ``Tunguska event").
The cause of this and other such events has remained unclear
and is the source of many debates (with frequent
conferences)\cite{tung}.
It is certainly remarkable that the fireball which
has been presumed to be an ordinary asteroid or comet 
simply disappears without trace in these events.
Indeed the strange properties of these events has lead
to purely geophysical explanations where it is proposed that
Tunguska and other similar events are produced by some
poorly understood coupling between tectonic and atmospheric
process\cite{olk}.  A fascinating possibility
is that these strange events are simply the manifestations
of the random collisions of the Earth with a mirror space body
as described above. 

Besides the purely scientific implications of this idea,
there is also another ramification:
If these strange events are due to mirror space bodies then it
may be difficult to protect the 
Earth against the threat of impact with these objects (which
may potentially pose an overall greater risk than space bodies
composed of ordinary matter).
An approaching space body made of (pure) mirror matter
would not be detectable (only after they impact
with the atmosphere would their
effects be observable, but then it would probably 
be to late to do anything about them).
However, mirror space bodies should contain some embedded ordinary
matter, whether or not it is enough for the space body
to be observable on its approach to Earth may be an
important issue if we want to try to prevent potentially
dangerous collsions.

Finally we finish with a few more related and hopefully 
interesting speculations. A popular theory for the origin 
of the moon is that it was formed when a large
large asteroid (or small planet) impacted
with the Earth during the early stages of the Earth's
formation. However, one of the problems
with this idea is that the chemical composition
of the moon is too similar to the Earth's mantle. There should
be a significant amount of extra-terrestrial material left over 
in the moon making the chemical composition
of the moon more different to that of the Earth's mantle
than it is known to be. However, if the colliding space body
was made of mirror
matter than this would alleviate this problem.
First, a smaller body may be needed if it was
made of mirror matter, especially if
the bodies kinetic energy is released below the surface 
because this should make
it easilier to liberate enough material
to form the moon. Second, any mirror material left on the
moon would eventually diffuse toward the moon's center.
In any case it would be undetectable and the 
composition of the moon would then appear similar
to that of the Earth's mantle. 
It is also possible that such collisions could help
explain the observed tilts in the axis of the sun and planets,
especially as mirror space bodies may 
orbit the sun in planes other than the ecliptic.

\newpage
\vskip 0.2cm
\noindent
{\bf Conclusion}
\vskip 0.2cm

It is a known fact that almost every 
plausible symmetry (such as rotational invariance, 
translational invariance etc) are found to be symmetries of
the particle interactions. Thus, it would be very strange if the
fundamental interactions were not left-right symmetric.
It is a very interesting observation that
left-right symmetry requires the existence of a
new form of matter called ``mirror matter" otherwise
there is nothing to balance the left-handed nature
of the weak force.  Even more interesting, is the 
remarkable evidence that mirror matter actually exists. 
The evidence ranges from studies of the most weakly
interacting elementary particles (the neutrinos) to 
evidence that most of the mass in the Universe is invisible
(i.e. dark matter). In fact seven fascinating puzzles have been
identified, each suggesting the existence of mirror matter.
While each individual puzzle is by itself not completely compelling, the
totality of the evidence is impressive.
Obviously if mirror matter does exist, it doesn't necessarily mean 
that all of the seven puzzles are
manifestations of the mirror world (although they may be).
Only further experiments/observations will provide the answer.
Nevertheless, the question of the existence of the mirror world
is one of the most interesting question in science
at the moment, and it should (hopefully) be answered within the 
next 5 years.

\vskip 0.5cm
\noindent
{\bf Acknowledgements}
\vskip 0.3cm
\noindent
I would like to thank Sergei Gninenko, Sasha Ignatiev, 
Henry Lew, Zurab Silagadze and 
especially Ray Volkas for collaboration on many of these ideas.
It is a pleasure to thank Andrei Ol'khovatov 
for correspondence.


\begin{thebibliography}{99}
\bibitem{bor}
Borexino Homepage: almime.mi.infn.it
\end{thebibliography}
\end{document}